\documentclass[11pt]{article}

\usepackage{a4wide,amssymb}
\newtheorem{theorem}{Theorem}[section]    
\newtheorem{lemma}[theorem]{Lemma}      
\newtheorem{corollary}[theorem]{Corollary}    
\newtheorem{definition}[theorem]{Definition} 
\newtheorem{fact}[theorem]{Fact}    
\newtheorem{result}[theorem]{Result}    
\newcommand{\qed}{\hfill{$\rule{6pt}{6pt}$}} 
\newenvironment{proof}{\noindent{\bf Proof}:}{\qed}

\newcommand{\ket}[1]{| #1 \rangle}
\newcommand{\bra}[1]{\langle #1 |}
\newcommand{\ketbra}[1]{| #1 \rangle \langle #1 |}

\newcommand{\Tr}{{\mathsf{Tr}}}

\newcommand{\defeq}{\stackrel{\Delta}{=}}

\newcommand{\EPR}{{\bf EPR}}
\newcommand{\PQC}{{\bf PQC}}\newcommand{\PQCE}{{\bf PQCE}}
\newcommand{\MPQC}{{\bf MPQC}}
\newcommand{\MPQCE}{{\bf MPQCE}}

\newcommand{\cC}{{\cal C}}
\newcommand{\cA}{{\cal A}}
\newcommand{\cB}{{\cal B}}

\newcommand{\cK}{{\cal K}}
\newcommand{\cH}{{\cal H}}
\newcommand{\cE}{{\cal E}}
\newcommand{\cD}{{\cal D}}

\title{Resource requirements of private quantum channels and
consequence for oblivious remote state preparation}
\author{
Rahul Jain \\
U.C. Berkeley\thanks{ Computer Science Division, University of California, Berkeley, CA~94720, USA.
Email: {\sf rahulj@cs.berkeley.edu}. Tel: +1-510-207-4102. Fax: +1-510-642-5775. This work was supported by an Army Research Office (ARO), North
California,  grant number DAAD 19-03-1-00082.
}
}
\date{}

\begin{document}
\maketitle
\begin{abstract}
Shannon~\cite{shannon:comm, shannon:sec} in celebrated works had shown
that $n$ bits of shared key is necessary and sufficient to transmit
$n$-bit classical information in an information-theoretically secure
way. Ambainis, Mosca, Tapp and de Wolf in~\cite{ronald:PQC} considered
a more general setting, referred to as {\em Private quantum channels},
in which instead of classical information, quantum states are required
to be transmitted and only one-way communication is allowed. They show
that in this case $2n$ bits of shared key is necessary and sufficient
to transmit an $n$-qubit state. We consider the most general setting
in which we allow for all possible combinations i.e. we let the input
to be transmitted, the message sent and the shared resources to be
classical/quantum. We develop a general framework by which we are able
to show simultaneously tight bounds on communication/shared resources
in all of these cases and this includes the results of Shannon and
Ambainis et al. 

As a consequence of our arguments we also show that in a one-way {\em
oblivious Remote state preparation} protocol for transferring an
$n$-qubit pure state, the entropy of the communication must be $2n$
and the {\em entanglement measure} of the shared resource must be
$n$. This generalizes on the result of Leung and
Shor~\cite{debby:obrsp} which shows same bound on the length of
communication in the special case when the shared resource is {\em
maximally entangled} e.g. $\EPR$ pairs and hence settles an open
question asked in their paper regarding protocols without maximally
entangled shared resource.
\\ \\
\noindent
{\em Key words: privacy, quantum channels, entropy, strong
sub-additivity, remote state preparation, substate theorem. }  
\end{abstract}

\section{Introduction}


Suppose Alice is required to transmit an $n$-bit input string to Bob
in an {\em information theoretically secure} way, i.e. without leaking
{\em any information} about her input to an eavesdropper Eve who has
complete access to the channel between her and Bob. Shannon
in~\cite{shannon:comm, shannon:sec} had shown that using $n$ bits of
shared key and by using {\em one-time pad} scheme Alice and Bob can
accomplish this. He further showed that $n$ bits of shared key are
also required by any other scheme which accomplishes the
same. Ambainis, Mosca, Tapp and de Wolf~\cite{ronald:PQC} considered a
generalization of this question in which instead of classical input,
Alice has quantum input and only one way of quantum communication between
Alice to Bob is allowed. They referred to this setting as Private
quantum channels ($\PQC$s). They showed that in this case the
requirement of shared key increases. Their main result was:
\begin{theorem}
$2n$ bits of shared key are necessary and sufficient to transmit any
$n$-qubit quantum state in an information-theoretically secure way.
\end{theorem}

We further generalize the setting by letting the shared resource
between Alice and Bob to be quantum.  A natural generalization of
classical shared keys in the context of quantum communication
protocols is a {\em pure} quantum state $\ket{\psi}^{AB}$ shared
between Alice and Bob.  This is referred to as {\em shared
entanglement} or simply entanglement. We consider private quantum
channels that use entanglement between Alice and Bob to achieve
security, and in order to distinguish them from $\PQC$s which use
classical shared keys, we call them $\PQCE$s. We formally define a
$\PQCE$ as follows. 
\begin{definition}
Let $S$ be a subset of pure $n$-qubit states.  Let $\ket{\psi}^{AB}$
be a bi-partite pure state shared between Alice and Bob and let $\rho$
be a quantum state.
\begin{enumerate}
\item {\bf Alice's operations:}
Alice gets an input pure state $\ket{\phi} \in S$. Alice's operation
consists of attaching a few ancilla qubits in the state $\ket{0}$ to
her input and her part of the bi-partite state $\ket{\psi}^{AB}$. She
then performs a unitary transformation on the combined quantum system
of all her qubits and sends a subset of the resulting qubits to
Bob. Let $\cA$ represent Alice's operations. Let for the input
$\ket{\phi}, \cE(\ket{\phi})$ represent the (encoded) quantum state of the qubits
sent to Bob. We have the following {\em
security requirement} that $\forall
\ket{\phi} \in S, \cE(\ket{\phi}) = \rho$.
\item {\bf Bob's operations:}
Bob on receiving the quantum message from Alice attaches a few ancilla
qubits in the state $\ket{0}$ to the combined system of the received
message and his part of the bi-partite state $\ket{\psi}^{AB}$. He
then performs a unitary transformation on the combined system of all
her qubits and outputs a subset of the resulting qubits. Let $\cB$
represent Bob's operations. Let for input state $\ket{\phi}$ to Alice
the final (decoded) output of Bob be represented by
$\cD(\ket{\phi})$. We have the following {\em correctness requirement}
that $\forall \ket{\phi}
\in S, \cD (\ket{\phi}) = \ketbra{\phi}$.

Then $[S,\cA, \cB, \ket{\psi}^{AB}, \rho]$ is called
a private quantum channel with entanglement ($\PQCE$).

\end{enumerate}
\end{definition}

\noindent  
{\bf Note:} 
\begin{enumerate}
\item From our description the mapping $\cE: \ket{\phi} \mapsto
\cE(\ket{\phi})$ (and extended by linearity to mixed states) from
Alice's input to her message, forms a {\em quantum operation}
(see Section~\ref{sec:prelim} for definition)
since it is a composition of quantum operations, like attaching
a fixed ancilla, performing unitary transformation and tracing out a
subsystem.
\item
In the above definition of a $\PQCE$, if we replace the
bi-partite shared pure state $\ket{\psi}^{AB}$ with shared random strings
between Alice and Bob, we get a $\PQC$. We represent a $\PQC$ by
$[S,\cA, \cB, P, \rho]$, where $P$ is the distribution of the shared
random strings between Alice and Bob.
\item In~\cite{ronald:PQC} the authors have made a comment that in the case of $\PQC$'s, without loss
of generality, Alice's operations can be thought of as the
following. On receiving the input she attaches a fixed mixed state
ancilla $\rho$ to it, applies a unitary $U_i$ depending on the shared
random string $i$ on the combined system of the input and the ancilla
and sends the resulting qubits to Bob. {\em \bf Please note that we do
not make such an assumption here which in any case does not apply for
$\PQCE$'s}. Also it is clear from the above definition that for both
$\PQC$'s and $\PQCE$'s, the operations of Alice and Bob are as general
as possible.

\item A $\PQCE/\PQC$ for $S$ is also $\PQCE/\PQC$
respectively for $\tilde{S}$ which is the closure of $S$ under convex combinations. 
\end{enumerate}

$\PQCE$s were also considered by Leung~\cite{debby:pqc} by the name of
{\em Quantum Vernam Cipher} who considered issues like security of key
recycling and reliability of message transfer. In this paper we are
primarily concerned with bounds on communication and entanglement
requirements of $\PQCE$s.  We consider the following measures of our
various resources:
\begin{definition}
\begin{itemize}
\item
{\bf Measure of communication:} For a $\PQC$ $[S,\cA, \cB, P,
\rho]$ and a $\PQCE$ $[S,\cA, \cB, \ket{\psi}^{AB}, \rho]$, we let the
measure of communication to be $S(\rho)$. When we say that
it requires '$n$ (qu)bits of communication' we mean $S(\rho)= n$. 
\item {\bf Measure of entanglement:}
For a bi-partite pure state $\ket{\psi}^{AB}$, consider its {\em
Schmidt decomposition}, $\ket{\psi}^{AB} = \sum_{i=1}^{k}
\sqrt{\lambda_i} \ket{a_i} \otimes \ket{b_i}$, where $\{\ket{a_i}\}$
is an orthonormal set and so is $\{\ket{b_i}\}$, $\lambda_i \geq 0$
and $\sum_i \lambda_i = 1$.  The measure of entanglement of
$\ket{\psi}^{AB}$ is defined to be $E(\ket{\psi}^{A B}) \defeq -\sum_i
\lambda_i \log \lambda_i$. For a $\PQCE$ $[S,\cA, \cB,
\ket{\psi}^{AB}, \rho]$, we let the measure of entanglement to be  $E(\ket{\psi}^{AB})$. When we say that it
requires $n$ ebits of entanglement we mean  $E(\ket{\psi}^{AB}) = n$.
\item {\bf Measure of shared randomness:} For a $\PQC$ $[S,\cA, \cB, P, \rho]$, we let the measure of
shared randomness be $S(P)$. When we say that it requires
$n$ bits of shared randomness we mean $S(P) = n$.
\end{itemize}

\end{definition}

We consider all possible cases i.e. when the input to Alice, the
message sent by Alice and the shared resource between Alice and Bob is
either classical or quantum. We develop a general argument by which we
are able to show tight bounds simultaneously on communication and
shared resource usage in all the above cases. Following is a compilation of
all the results we obtain due to our analysis. Below when we say the
``x,y,z'' case (e.g. classical, quantum, classical case) we mean,
Alice gets $n$-(qu)bits of x input, the communication is y and the
shared resource is z.

\begin{result}
\label{result:main}
\begin{enumerate}
\item In the classical, classical, classical case, $n$ bits of
communication and $n$ bits of shared key is 
required. The one-time pad scheme hence is simultaneously optimal in
both communication and shared key usage. This is basically Shannon's
result~\cite{shannon:comm, shannon:sec}. 

\item In the classical, quantum, classical case, $n$
qubits of communication  and $n$ bits of shared key is
required. Hence here again simultaneously optimal upper bound is
achieved by the one-time pad scheme. 

\item In the classical, classical, quantum case, $n$
bits of communication and $n$ ebits of entanglement is
required. Hence here again simultaneously optimal upper bound is
achieved by the one-time pad scheme.

\item In the classical, quantum, quantum case, $n/2$
qubits of communication and $n/2$ ebits of entanglement is required.
The simultaneously optimal upper bound here is achieved by the
standard protocol for {\em super-dense
coding}~\cite{bennett:superdense, nielsen:quant} which is a
$\PQCE$. In it, Alice transfers $n$ bits of classical input in an
information-theoretically secure manner to Bob using $n/2$ qubits of
communication and $n/2-\EPR$ pairs~\cite{nielsen:quant} shared between
them. In this case the message of Alice is always in the maximally
mixed state independent of her input.

\item The quantum, classical, classical case is impossible with finite
communication.

\item In the quantum, quantum, classical case, $n$ qubits of
communication and $2n$ bits of shared key is required. This is the
main result of Ambainis et al.~\cite{ronald:PQC}.  In the same paper they have
exhibited a $\PQC$ which transfers an $n$-qubit state with $n$-qubits
of communication and $2n$ bits of shared randomness and is therefore
simultaneously optimal in both communication and shared randomness.

\item In the quantum, classical, quantum case, $2n$ bits of
communication and $n$ ebits of entanglement is required. Here the
simultaneously optimal scheme is the standard protocol for {\em
teleportation}~\cite{bennett:teleport, nielsen:quant} which is a
$\PQCE$ (pointed to us by de Wolf in personal communication). In this
protocol Alice can transfer $n$-qubits to Bob in an information
theoretically secure way by using $2n$ bits of communication and using
$n-\EPR$ pairs between them. In this case the message of Alice always
has uniform distribution independent of her input.

\item  In the quantum, quantum, quantum case, $n$ qubits of communication
and $n$ ebits of entanglement is required. In this case simultaneously
optimal upper bound is achieved by a scheme using (2,3)
quantum secret sharing scheme by Cleve, Gottesman and Lo~\cite{cleve:secret}.
(This scheme was pointed by to us by Nayak who in turn was pointed to
by Gottesman).  

\end{enumerate}

\end{result}

We also present a consequence of our results to one-way, oblivious,
remote state preparation (RSP) protocols. In an RSP protocol between
Alice and Bob, Alice is required to transport a known quantum state
$\ket{\phi}$ in $n$-qubits to Bob using classical communication and
some shared entanglement. An RSP is called oblivious if at the end of
the protocol, Bob gets a copy of Alice's input $\ket{\phi}$ and rest
of his qubits are independent of $\ket{\phi}$.  Leung and
Shor~\cite{debby:obrsp} have shown that for one-way oblivious RSPs, if
Alice and Bob start with maximally entangled state then the worst case
communication required by them is $2n$. We generalize on their result
to provide bounds for all one-way oblivious RSP protocols independent
of which shared pure state they start with. We prove that for any
one-way oblivious RSP protocol, the entropy of communication is at
least $2n$ and the entanglement measure of the shared pure state is at
least $n$. Therefore again teleportation achieves both these bounds
simultaneously.

Finally we discuss two-way multiple round $\PQC$s ($\MPQC$s) and
$\PQCE$s ($\MPQCE$s). We show that an $\MPQC$ which can transfer an
$n$-qubit state must use $n$-bits of classical shared keys. Also an
$\MPQCE$ which can transfer an $n$-qubit state must use $\Omega(n)$
ebits of entanglement. Hence there is not much saving even when
multiple rounds are allowed.

\section{Organization of the paper}
In the next section we make a few definitions and state a few facts
which we will be using later in our proofs. In
Section~\ref{sec:bounds} we present the proofs of all the parts of
Result~\ref{result:main}. In subsection~\ref{subsec:rsp} we discuss our
result for one-way oblivious RSPs. In Section~\ref{sec:multi} we
discuss two-way multiple round private quantum channels and finally
conclude with a few remarks in Section~\ref{sec:conclude}.

\section{Preliminaries}
\label{sec:prelim}
Let $\cH_{k}$ represent the Hilbert space of dimension $k$. Let
$\cC_k$ represent the set of quantum states corresponding to the
standard basis of $\cH_{k}$, also referred to as the {\em classical
states}. Let $I_{k}$ represent the identity transformation in a $k$
dimensional space. For an operator $A$ let $A \geq 0$ represent that
$A$ is a positive semi-definite operator. By a quantum operation
we mean a linear, completely positive, trace-preserving operation. Let
$\cH, \cK$ be Hilbert spaces. For a state $\rho \in \cK$, we call a
pure state $\ket{\phi} \in \cH
\otimes \cK$, a {\em purification} of $\rho$ if $\Tr_{\cH}
\ketbra{\phi} = \rho$. Let us represent the four
{\em Pauli operators} in the standard basis as $\sigma_0 \defeq \left(
\begin{array}{cc} 1 & 0 \\ 0 & 1 \end{array} \right ), \sigma_1 \defeq
\left( \begin{array}{cc} 0 & 1 \\ 1 & 0 \end{array} \right ), \sigma_2
\defeq \left( \begin{array}{cc} 0 & i \\ -i & 0 \end{array} \right ),
\sigma_3 \defeq \left( \begin{array}{cc} 1 & 0 \\ 0 & -1 \end{array}
\right ) $. Let us identify a state in $\cC_{2^{2n}}$ as a string
$x(\defeq x_1x_2 \ldots x_n) \in \{0,1,2,3\}^n$ in the natural way by
pairing up the bits from left to right. Let $\sigma_x \defeq
\sigma_{x_1} \otimes \sigma_{x_2} \ldots \otimes \sigma_{x_n}$. Let an
$\EPR$ pair mean the state $\ket{\EPR} \defeq \frac{1}{\sqrt{2}}(\ket{00} +
\ket{11})$. For  $s \in \{0,1,2,3\}$, the states $ (\sigma_s \otimes
I) \ket{\EPR}$ are referred to as the four {\em Bell states}. Please note
that all the four Bell states are orthogonal to each other. 

For a quantum state $\rho$ with eigenvalues $\lambda_i$ its {\em von-Neumann entropy} is defined as $S(\rho) \defeq -\sum_i
\lambda_i \log \lambda_i$. Given a joint quantum system $AB$, the
mutual information between them is defined as $I(A:B) \defeq S(A) +
S(B) - S(AB)$.  Relative entropy between two states $\rho$ and $\sigma$ is
defined as $S(\rho | \sigma ) \defeq \Tr \rho ( \log \rho - \log
\sigma )$.  We require the following properties of von-Neumann entropy,
relative entropy and mutual information. Please refer
to~\cite{nielsen:quant} for a good introduction to quantum information
theory.
\begin{fact}
\label{fact:prop}
\begin{enumerate}
\item
 $S(A) + S(B) - S(AB) \geq 0$.  This is called as sub-additivity
 property of von-Neumann entropy. This implies  $I(A:B) \geq 0$. 
\item $S(ABC) + S(A) \leq S(AB) + S(AC)$. This is called the strong
sub-additivity property. This implies $I(\cE(A):B) \leq I(A:B)$, where
$\cE$ is a quantum operation. 
\item 
We have the following chain rule of mutual-information, $I(A:BC) =
I(A:B) + I(AB:C) - I(B:C)$, which follows easily from definition.
\item 
$S(AB) \geq | S(A) - S(B) |$. This is called as {\em Araki-Lieb}
inequality.  
\item Given a bi-partite system $\rho^{AB}$, $I(A:B) = S(\rho^{AB} | \rho^A \otimes \rho^B)$, where $\rho^A, \rho^B$ are the states of the
systems  $A$ and $B$ respectively.
\item Given a joint system $AB$ with $A$ being a classical system,
$S(AB) \geq \max \{S(A), S(B)\}$. 

\end{enumerate}
\end{fact}

We will need the following theorem. 

\begin{theorem}[Local transition
theorem~\cite{mayer:imp,hklo:bitcomm1, hklo:bitcomm2}] 
\label{thm:local} 
Let $\cK, \cH$ be Hilbert spaces. Let $\rho$ be a quantum state in
$\cK$. Let $\ket{\phi_1}$ and $\ket{\phi_2}$ be two purification of
$\rho$ in $\cH \otimes \cK$. Then there is a local unitary
transformation $U$ acting on $\cH$ such that $(U \otimes I)
\ket{\phi_1} = \ket{\phi_2}$.
\end{theorem}

We will also need the following {\em Substate} theorem
from~\cite{jain:substate}.
\begin{fact}
Let $\rho, \sigma$ be quantum state. If $S(\rho | \sigma)\leq k$
then,
$$ \sigma - \frac{\rho'}{2^{O(k)}}   \geq 0$$ where 
$\Tr |\rho' - \rho | \leq 0.1 $.
\end{fact}
 

\section{Resource  bounds}
\label{sec:bounds}
We first derive a few lemmas which will finally lead us to our
results. In~\cite{ronald:PQC} it is shown that a $\PQC$ which can
transmit $n$-qubit quantum states can be converted into a $\PQC$ which
uses the same amount of shared classical randomness to transmit any
$2n$ bit classical state. We show a similar thing for $\PQCE$'s. Following lemma states the same.

\begin{lemma}
\label{lem:convert}
If there exists a $\PQCE$, $[\cH_{2^n}, \cA, \cB, \ket{\psi^{AB}},
\rho]$ then there exists a $\PQCE$, \\ $[\cC_{2^{2n}},\cA', \cB',
\ket{\psi^{AB}}, I_{2^n} \otimes \rho ]$ which uses the same bi-partite
state as the shared entanglement between Alice and Bob and uses extra
$n$-qubits of communication.
\end{lemma}

In order to prove this lemma we first prove here another lemma which
is very similar to a lemma from~\cite{ronald:PQC}.
\begin{lemma}
Let $\cH, \cK$ be Hilbert spaces. Let $\cE$ be a quantum
operation acting on $\cH$ such that $\forall \ket{\phi}
\in \cH, \cE(\ketbra{\phi}) = \rho $.
Let $\ket{\phi_1}, \ket{\phi_2} \in \cH$ be two
orthogonal states, then $\cE(\ket{\phi_1}\bra{\phi_2}) =
\cE(\ket{\phi_2}\bra{\phi_1})= 0$. 
\end{lemma}
\begin{proof}
We note the following:
\begin{eqnarray}
\rho & = & \cE(\ketbra{\phi_1}) = \cE(\ketbra{\phi_2}) \\
\rho & = & \cE(\frac{1}{2} (\ket{\phi_1} + \ket{\phi_2}) (\bra{\phi_1} +
\bra{\phi_2})  \\
\rho & = & \cE(\frac{1}{2} (\ket{\phi_1} + i \ket{\phi_2}) (\bra{\phi_1} -i \bra{\phi_2}) 
\end{eqnarray}
Now (1) and (2) imply $\cE(\ket{\phi_1}\bra{\phi_2}) +
\cE(\ket{\phi_2}\bra{\phi_1}) = 0 $ and (1) and (3) imply $\cE(\ket{\phi_1}\bra{\phi_2}) -
\cE(\ket{\phi_2}\bra{\phi_1}) = 0 $. Together the two imply
$\cE(\ket{\phi_1}\bra{\phi_2}) = \cE(\ket{\phi_2}\bra{\phi_1}) = 0$.
\end{proof}
\\ \\
We get the following corollary of the above lemma:
\begin{corollary}
\label{cor:nice}
Let $\cH, \cK$ be Hilbert spaces. Let $\cE$ be a quantum operation acting on $\cH$ such that $\forall \ket{\phi}
\in \cH, \cE(\ketbra{\phi}) = \rho $. Then $\forall \ket{\psi} \in \cK
\otimes \cH, (I \otimes \cE) (\ketbra{\psi}) = (\Tr_{\cH}
\ketbra{\psi}) \otimes \rho$. This also means that for all mixed states
$\sigma \in \cK \otimes \cH, (I \otimes \cE) \sigma = (\Tr_{\cH} \sigma) \otimes \rho$.     
\end{corollary}

\begin{proof}
Let $\ket{\psi} = \sum_i \sqrt{\lambda_i} \ket{a_i} \otimes \ket{b_i}$, be as
written in the Schmidt decomposition form. Then,
\begin{eqnarray*}
(I \otimes \cE) (\ketbra{\psi}) & = &  (I \otimes \cE) (\sum_i \sqrt{\lambda_i}
\ket{a_i} \otimes \ket{b_i})(\sum_j \sqrt{\lambda_j} \bra{a_j} \otimes \bra{b_j}) \\
& = & \sum_{i,j} (I \otimes \cE) \sqrt{\lambda_i}  \sqrt{\lambda_j}
\ket{a_i}  \bra{a_j} \otimes  \ket{b_i} \bra{b_j}  \\
& = & \sum_{i,j}   \sqrt{\lambda_i}  \sqrt{\lambda_j}
\ket{a_i}  \bra{a_j} \otimes  \cE(\ket{b_i} \bra{b_j}) \\
& = & \sum_{i}   \lambda_i \ketbra{a_i} \otimes  \cE (\ketbra{b_i}) \\
& = & (\sum_{i}   \lambda_i \ketbra{a_i}) \otimes \rho \\
& = & \Tr_{\cH} \ketbra{\psi}) \otimes \rho
\end{eqnarray*}

\end{proof}

\begin{proof}{\bf (Lemma~\ref{lem:convert}) }
In the $\PQCE, [\cC_{2^{2n}},\cA', \cB', \ket{\psi^{AB}}, I_{2^n}
\otimes \rho ]$, let $x \in \{0,1,2,3 \}^n$ correspond to the input state. Alice prepares $n~\EPR$ pairs and applies the unitary
$\sigma_x$ on combined system of the first qubits of each pair. She
then encrypts the combined system of the second qubits of each pair
using $\cE$, the encryption operation of the $\PQCE, [\cH_{2^n}, \cA,
\cB,
\ket{\psi^{AB}},
\rho]$. She now sends all the resulting qubits to Bob. From above corollary, we can see that the state of the
message of this new $\PQCE$ will be $I_{2^n} \otimes \rho$ for all
inputs in $\cC_{2^{2n}}$.
The decryption operation $\cB'$ of Bob now corresponds to
first decrypting the second half of the received qubits using $\cB$
and then recovering the input classical state by making measurements on
the $n$-Bell states.
\end{proof}

Below we show a similar lemma which implies that a $\PQC/\PQCE$ which
transmits any $n$-qubit quantum state can be converted into a $\PQC/\PQCE$
which uses the same communication and extra $n$ ebits of entanglement
to transmit any $2n$ bit classical state. We show the proof for
$\PQCE$s and a similar proof holds for $\PQC$s as well.

\begin{lemma}
\label{lem:convert2}
If there exists a $\PQCE$, $[\cH_{2^n}, \cA, \cB, \ket{\psi^{AB}},
\rho]$ then there exists a $\PQCE$, \\ $[\cC_{2^{2n}},\cA', \cB',
 \ket{\psi^{AB}} \otimes (\frac{\ket{00} +
\ket{11}}{\sqrt{2}})^{\otimes n}, \rho ]$ which uses the same communication
and extra $n$-{\EPR} pairs. 
\end{lemma}

\begin{proof}
In $[\cC_{2^{2n}},\cA', \cB',  \ket{\psi^{AB}} \otimes (\frac{\ket{00} +
\ket{11}}{\sqrt{2}})^{\otimes n}, \rho ]$, let $x \in \{0,1,2,3 \}^n$
correspond to the input state. Alice applies $\sigma_x$ to her part of
the extra $n$-$\EPR$ pairs, encodes them using the encoding procedure
of the earlier $\PQCE$ $[\cH_{2^n}, \cA, \cB, \ket{\psi^{AB}}, \rho]$, and sends the resulting
qubits to Bob. The security property of $[\cH_{2^n},
\cA, \cB, \ket{\psi^{AB}}, \rho]$ implies the security property of
$[\cC_{2^{2n}},\cA', \cB', \ket{\psi^{AB}} \otimes (\frac{\ket{00} +
\ket{11}}{\sqrt{2}})^{\otimes n}, \rho ]$. On receiving
Alice's message, Bob first applies the decoding procedure of
$[\cH_{2^n}, \cA, \cB, \ket{\psi^{AB}}, \rho]$, and recovers $x$ by
making measurements on the $n$-Bell states.
\end{proof}

We will need the following lemma.
\begin{lemma}
\label{lem:basic}
Let $ABX$ be a tripartite system. Then,
\begin{enumerate}
\item 
$S(AX)+ S(BX) - S(ABX) - S(X) \leq \min \{2S(A), 2S(B)\}$. 
\item 
If $AX$ is a classical system then we have the stronger inequality
$S(AX)+ S(BX) - S(ABX) - S(X) \leq \min \{S(A), S(B)\}$.
\item $I(A:B) \leq \min \{2S(A), 2S(B)\}$.
\end{enumerate}
\end{lemma}
\begin{proof}
\begin{enumerate}
\item
\begin{eqnarray*}
S(AX) - S(ABX) + S(BX) - S(X) & \leq & S(AX) - S(ABX) + S(B) \\ & \leq
& S(B) + S(B) = 2 S(B)
\end{eqnarray*}
Above first inequality comes from part (1) and
second  inequality comes from part (4) of
Fact~\ref{fact:prop}. Similarly we get $S(AX)+
S(BX) - S(ABX) - S(X) \leq 2 S(A) $. 
\item
$$S(AX) - S(ABX) + S(BX)  - S(X)  \leq  S(BX) - S(X) \leq  S(B) $$

Above first inequality arises from part (6), since $AX$ is a classical
system, and the second inequality comes from part (1) of
Fact~\ref{fact:prop}. Again, since $A$ is a classical
system, we get 
$$S(AX) - S(X)  + S(BX) - S(ABX) \leq  S(AX) - S(X) \leq S(A) $$
Above the first inequality comes from part (6) and the second
inequality comes from part (1) of Fact~\ref{fact:prop}.
\item 
$$ I(A:B) = S(A) + S(B) - S(AB) \leq S(A) + S(A) = 2S(A) $$
The inequality above follows from part (4) of Fact~\ref{fact:prop}. 
\end{enumerate}
\end{proof}

We now have the following theorem.
\begin{theorem}
\label{thm:main}
If $[\cC_{2^n},\cA, \cB, \ket{\psi^{AB}}, \rho ]$ is a $\PQCE$ then,
\begin{enumerate}
\item $S(\sigma^B) \geq n/2$, where $\sigma^B$ is the quantum state
corresponding to Bob's part of $\ket{\psi^{AB}}$. We note from
definitions that $S(\sigma^B) = E(\ket{\psi}^{AB})$.
\item  $S(\rho) \geq n/2$.
\end{enumerate}
\end{theorem}

\begin{proof}
Let $X$ be a random variable which takes values in $\{1,2, ....,
2^n\}$ uniformly and through the $\PQCE$ Alice is able to communicate
$X$ to Bob.  We can assume that the operations of Alice are {\em safe}
on $X$ which means that at the beginning Alice makes a copy of $X$
(since it is a classical state) and then her subsequent operations
do not touch the original copy of $X$. Let $M_1$ be the quantum state
corresponding to the message of Alice and let $M_2$ be the quantum
state corresponding to Bob's part of $\ket{\psi}^{AB}$. Then from
Fact~\ref{fact:prop},
\begin{eqnarray*}
n & = & H(X) = I(\cD(X) : X)  =  I(\cB(M_1M_2 \otimes \ketbra{0}_{ancilla}) : X) \\ 
&\leq& I(M_1M_2\otimes \ketbra{0}_{ancilla} : X) =  I(M_1M_2 : X) \\
& =&  I(M_1 : X) + I(M_2 : M_1X)  - I(M_1 : M_2) \\ 
& = &  I(M_1 : X) + I(M_2 : X) + I(M_2X : M_1) - I(M_1:X) - I(M_1 : M_2) \\ 
& \leq & 0 + 0 + I(M_2X : M_1) - I(M_1:X)  \\ 
& = & S(M_2X) + S(M_1X) - S(M_1M_2X) - S(X) \\ 
& \leq & \min\{2S(M_2), 2S(M_1)\}
\end{eqnarray*}
Above, first inequality comes from part (2) of Fact~\ref{fact:prop}.
$I(M_1 : X) = 0$ because of the privacy property of the channel.
$I(M_2 : X) = 0$ because they were independent to begin with and
Alice's operations are safe on $X$. The last inequality follows from
part (1) of Lemma~\ref{lem:basic}.
\end{proof}

We note in the proof of Theorem~\ref{thm:main}, due to part (2) of
Lemma~\ref{lem:basic}, that if either $M_2$ is a classical system (as
in a $\PQC$) or if $M_1$ is a classical system (when the message is classical), then we get $n \leq
\min \{S(M_2), S(M_1) \}$. Therefore we have the following corollary:

\begin{corollary}
\label{cor:class}
\begin{enumerate}
\item 
If $[\cC_{2^n},\cA, \cB, P, \rho ]$ is a $\PQC$ then, $S(P) \geq n$ and
$S(\rho) \geq n$.  
\item 
If $[\cC_{2^n},\cA, \cB, \ket{\psi^{AB}}, P ]$
is a $\PQCE$ with classical communication then,
$S(\sigma) \geq n$, where $\sigma$ is Bob's part of $\ket{\psi^{AB}}$,
and $S(P) \geq n$. 
\end{enumerate}
\end{corollary}

We are now set to show various parts of Result~\ref{result:main}.

\begin{proof} \begin{enumerate} \item Follows from part (1) of
Corollary~\ref{cor:class}.  \item Follows from part (1) of
Corollary~\ref{cor:class}.  \item Follows from part (2) of
Corollary~\ref{cor:class}.  \item Follows from Theorem~\ref{thm:main}.
\item Follows from the fact that quantum states cannot be encoded as
finite classical distributions and faithfully recovered.  \item
\label{it:pqc} From $\PQC$ $[\cH_{2^n},\cA, \cB, P, \rho]$, using
Lemma~\ref{lem:convert} we get a $\PQC$ $[\cC_{2^{2n}},\cA', \cB', P,
I_{2^n} \otimes \rho]$. Part (1) of Corollary~\ref{cor:class} now
implies $S(P) \geq 2n$.  Lower bound on communication follows from the
fact that a $\PQC$ for $\cH_{2^n}$ is also a $\PQC$ for $\cC_{2^{n}}$
and Part (1) of Corollary~\ref{cor:class}.

\item \label{it:pqce} Lower bound on communication follows from
Lemma~\ref{lem:convert2} and Part (2) of
Corollary~\ref{cor:class}. Lower bound on entanglement
follows from the fact that a $\PQCE$ for $\cH_{2^n}$ is also a $\PQCE$
for $\cC_{2^{n}}$ and Part (2) Corollary~\ref{cor:class}.  

\item From
$[\cH_{2^n},\cA, \cB, \ket{\psi^{AB}}, \rho]$ using
Lemma~\ref{lem:convert} we get a $\PQCE$ $[\cC_{2^{2n}},\cA', \cB',
\ket{\psi^{AB}}, I_{2^n} \otimes \rho]$. Theorem~\ref{thm:main} now
implies $E(\ket{\psi^{AB}}) \geq n$. Similarly lower bound on
communication follows from the Lemma~\ref{lem:convert2} and
Theorem~\ref{thm:main}.  

\end{enumerate} 

\end{proof}

\subsection{Consequence for one-way oblivious remote state preparation
problem} \
\label{subsec:rsp}In a remote state preparation (RSP) protocol between Alice
and Bob, Alice wants to transport a known $n$-qubit pure state
$\ket{\phi}$ to Bob using classical communication and shared prior
entanglement. Such a protocol is called oblivious if at the end of the
protocol, Bob gets a single copy of Alice's input $\ket{\phi}$ and
other than that all his qubits are independent of $\ket{\phi}$.

Let us consider an RSP protocol. Let $\rho$ be Bob's part of the
initial pure state shared between Alice and Bob. Let the state of the
shared part of entanglement on Bob's side after receiving message $m$
to be $\rho^{\ket{\phi}}_m$. Since the protocol is oblivious, the
probability with which a particular message $m$ comes to Bob is
independent of $\ket{\phi}$, which we denote by $p_m$.  Therefore we
note $\sum_m p_m \rho^{\ket{\phi}}_m = \rho$ for all $\ket{\phi}$
(since the entanglement part of Bob's qubits has not changed due to
Alice's operations).

Bob on receiving message $m$ attaches ancilla $\ket{0}$ to her qubits
and performs unitary $U_m$ to them. Again since the protocol is
oblivious, her state at the end of the unitary is $\ketbra{\phi}
\otimes \sigma_m$, where $\sigma_m$ is independent of
$\ket{\phi}$. Using these properties we now construct a $\PQC$ between
Alice and Bob. Let Alice and Bob share classical randomness between
them in which $m$ is appears with probability $p_m$. On shared string
being $m$, Alice attaches $\sigma_m$ to $\ketbra{\phi}$, applies
$U_m^{\dagger}$ and sends the resulting state $\rho^{\ket{\phi}}_m
\otimes \ketbra{0}$ to Bob. Now since  $\sum_m p_m \rho^{\ket{\phi}}_m
= \rho$ for all $\ket{\phi}$, Alice's message is independent of
$\ket{\phi}$ and hence the privacy requirement is satisfied. Bob
on receiving the quantum message applies $U_m$ to it and discards
$\sigma_m$. Therefore now from part 6 of Result~\ref{result:main} we get $S(\rho)
\geq n$ and $\sum_m - p_m \log p_m \geq 2n$.

\section{Multiple round private quantum channels}
\label{sec:multi}




When we consider two-way multiple round $\PQC$s (denoted $\MPQC$) or
multiple round $\PQCE$s ($\MPQCE$), we note that keeping the privacy
of {\em individual messages} cannot be the only criteria. For example
let us consider a protocol in which in the first message Alice
transfers $\EPR$ pairs followed by a junk message of Bob and then
Alice transfers her quantum state privately using the earlier sent
$\EPR$ pairs. In this protocol none of the individual messages give
any information about the transfered state but it does not mean that
Eve, who can access the channel in all rounds, cannot get any
information about the transfered state. 

We therefore consider two possible definitions of $\MPQC$s and
$\MPQCE$s. We define $\MPQC$s and $\MPQCE$s are similar with only
shared randomness replaced by shared entanglement.

\begin{enumerate}
\item 
{\bf $\MPQC$s without abort:} In this case Alice and Bob never abort
the protocol but satisfy the following:
\begin{itemize}
\item Any interfering Eve gets no information about the
 input state of Alice. 
\item If Eve is not interfering then  the input state is
faithfully transfered to Bob.
\end{itemize}

\item
{\bf $\MPQC$s with abort:} In this case Alice can abort the protocol
any time but satisfy the following:
\begin{itemize}
\item Before abort any interfering Eve gets no information about the
input state of Alice. 
\item If there is no abort then the input state is
faithfully transfered to Bob. 
\end{itemize}
\end{enumerate}
\noindent
{\bf Remark:} Consider an implementation of a private quantum channel
in which Alice and Bob first use {\em quantum key distribution} (QED)
protocols like BB84 for key generation and then use these keys to
transfer quantum states privately. However it is not strictly an
$\MPQC$ according to our definition, because current implementations
of QEDs require the existence of a classical broadcast channel which
is unjam able by Eve. Also such a protocol would not be perfectly
secure and there would still be a small amount of information that
Eve can obtain even in case Alice does not abort the protocol.


Below we discuss the resource requirements of $\MPQC$s and
$\MPQCE$s. The cheating strategies of Eve discussed below work
in both type of protocols, with and without abort.

\begin{lemma}
Let $P$ be the distribution of the shared random strings between
Alice and Bob in an $\MPQC$ for $\cC_{2^n}$. Then
$S(P) \geq n$.

\end{lemma}
\begin{proof}
Consider an attack of Eve where she starts acting like Bob. She
guesses the random string which has highest probability, say $p$ of
occurring.  The probability that her guessed string is equal to Alice's
random string is at least $p$. In the event that she guesses Alice's
random string correct, she gets to know Alice's input state faithfully
at the end of the protocol and Alice does not abort the protocol in
this case. Hence from the security criterion, $
p \leq 2^{-n}$. This implies $S(\sigma) \geq n$.
\end{proof}

We show a similar statement for  $\MPQCE$s.
\begin{lemma}
Let $\ket{\psi}^{AB}$ be the prior shared pure state between Alice and
Bob in an $\MPQCE$s for $\cC_{2^n}$. Let $\sigma^{AB} =
\ketbra{\psi}$. Let $\sigma^A$ and $\sigma^B$ denote state of Alice's
and Bob's parts respectively in $\sigma^{AB}$. Then
$E(\ket{\psi}^{AB})= S(\sigma^A) = S(\sigma^B) = \Omega(n)$.
\end{lemma}
\begin{proof}
Let $S(\sigma^B) = k$. Similar to above, let us consider a cheating
strategy of Eve in which she starts acting like Bob. She starts with
the state $\sigma^B$ in the register which holds Bob's part of the
entanglement. Let $M_1$ and $M_2$ represent Alice and Bob's parts in
$\sigma^{AB}$. Then, from Lemma~\ref{lem:basic} we get, 
$$ S(\sigma^{AB} | \sigma^A \otimes \sigma^B ) = I(M_1:M_2) \leq
2S(\sigma_B) = 2k $$ From substate theorem, $$ \sigma^A \otimes
\sigma^B - \frac{\sigma'^{AB}}{2^{O(k)}} \geq 0$$ where $\Tr
|\sigma'^{AB} - \sigma^{AB}| \leq 0.1$

This implies that Eve with probability $2^{-O(k)}$ gets the same state
created with her when Alice and Bob start with $\sigma'^{AB}$ as the
prior entangled state. Because $\Tr |\sigma'^{AB} - \sigma^{AB}| \leq
0.1$, Alice's probability of abort $\leq 0.1$. Hence the state created
with Eve will be the same as the input state of Alice
with probability at least $(0.8)2^{-O(k)}$. Because of the security
criterion $(0.8)2^{-O(k)} \leq 2^{-n} \Rightarrow k = \Omega(n)$.
\end{proof}

\section{Conclusion} \label{sec:conclude} We have considered private
quantum channels with one-way communication of all possible kinds and
in all the cases we have shown simultaneously optimal resource
requirements. Even when we allow two-way communication but if Eve is
allowed arbitrary access to the channel, we show that there is not much
saving possible on prior entanglement/shared randomness. However, by allowing a
classical broadcast channel between Alice and Bob, unjam able by Eve,
saving is possible on prior entanglement/shared randomness by using
QKD protocols. So is there a weaker assumption we can make for saving on
prior entanglement/shared randomness? 

In connection with RSPs it will be interesting to show similar bounds
on resources when we do not have the oblivious condition or for
two-way multiple round (non)-oblivious protocols. 
\\ \\
\noindent
{\large \bf Acknowledgment:} We thank Daniel Gottesman, Hartmut Klauck, Gatis
Midrijanis, Ashwin Nayak and Ronald de Wolf, for useful discussions. We thank Andris Ambainis for
pointing reference~\cite{debby:pqc} and Jaikumar Radhakrishnan and Pranab Sen for useful comments
on an earlier draft.

\newcommand{\etalchar}[1]{$^{#1}$}

\end{document}